\documentstyle[twocolumn,epsf,prl,aps]{revtex}
\usepackage[dvips]{graphicx}

\begin{document}
\hyphenation{Nij-me-gen}
\hyphenation{Shklyarevskii}
\title{Keijsers, Shklyarevskii and van Kempen Reply}
\author{R.J.P. Keijsers$^{1}$, O.I. Shklyarevskii$^{1,}$ $^{2}$,  
H. van Kempen$^{1}$}
\address{$^{1}$ Research Institute for Materials, University of 
Nijmegen,\\ Toernooiveld 1, 6525 ED, Nijmegen, The Netherlands}
\address{$^{2}$ B. Verkin Institute for Low  Temperature Physics \& 
Engineering,\\ National Academy of Science of Ukraine, 47 Lenin Av.,
310164, Kharkov, Ukraine}
\maketitle

\nopagebreak

\begin{abstract}
Answer to the Comment on ``Point-Contact Study of Fast and Slow Two-Level
Fluctuators in Metallic Glasses'' by Jan von Delft et al.
\end{abstract}

In our paper\cite{Keijsers95} we described a zero-bias anomaly (ZBA) obtained
in the point-contact spectra of  point-contacts of  mechanically distorted
metals.  The anomaly  could be ascribed to the influence of conduction electron
interaction with two level systems (TLS).  Different mechanisms,  based on
non-magnetic Kondo like interactions\cite{VlaZa} (VZ) and on elastic and
non-elastic electron-TLS scattering\cite{KozKul} (KK), can  explain the
results. However, on the basis of the experiment we could not decide which
mechanism is the most appropriate to explain the data. The situation is
specially complicated because the different mechanisms  are not mutually
exclusive but can be present simultaneous.  In a later paper\cite{Keijsers} we
described  a similar experiment, but now with metallic glasses for which a high
density of TLSs can be expected.  Indeed a strong ZBA is present.  In addition
it was observed that the system often switched between two (and sometimes more)
different ZBAs producing telegraph noise variations in the resistance.  This
switching we explained as being caused by the occurrence of configurational
changes in the neighborhood of the contact which influenced the TLSs.  Again te
results could be explained  within  the frame work of the VZ and KK models.

In the Comment \cite{vonDelft} von Delft et al. correctly state that the strong
resemblance of the VZ and KK predictions is to a large extend due to averaging
over ensembles of TLSs and that, when a single TLS could be observed, a better
comparison with the models can be made. They conclude from the smallness of the
differences between the ZBAs we find in ref. \cite{Keijsers} that only one (or
a few) TLS is involved and so the differences should show the characteristics
of a single (or a few) TLS. Comparing those differences with the theoretical
models they conclude that the VZ mechanism gives clearly the best fit (see
figure in ref. \cite{vonDelft}).  This is an interesting alternative way to
analyse our data and we can agree with the conclusion obtained that the VZ
model  is dominant.  However, we want to emphasize that the dominancy of the VZ
model does not exclude the KK mechanisms.  Generally speaking, in point-contact
spectroscopy every scattering on excitations has to show up although, of course,
with amplitudes which depend on many factors.  Indeed the comparison in fig. 1
of ref. \cite{vonDelft} shows that there is room for a contribution of the KK
elastic scattering mechanism. 

An interesting question is which other experiments could discriminate between
the different mechanisms. Von  Delft  et al.  \cite{vonDelft} mentioned a V/T
scaling analysis. An other experiment involves  the measurement of the
influence of the relaxation times which are very different for the different
processes. The relaxation  times can be deduced from RF response signals of the
point-contacts. This last experiment has been done by Balkashin et al. 
\cite{Balkashin} with the same materials  as used in ref.\cite{Keijsers}. The
results agree with the preceding analysis. The VZ mechanism is dominant, but
again the data do not exclude a contribution of the elastic KZ  mechanism.
The determination of the size of the last contribution however, is hampered by
the presence of a background signal, coming from electron-phonon and
electron-electron scattering, which is not known accurately enough to allow
subtraction.  

Another point we want to stress is that different materials can contain TLSs of
a different nature, with different electron-TLS interactions. For example the
ZBAs discussed in ref. 1 showed different signs, indicating that different TLSs
might have been observed. Of course all these remarks do not lessen the
importance of the basic observation of von Delft et al. that it is possible to
study single TLSs as described in their Comment.\\

\noindent
R.J.P. Keijsers$^{1}$, O.I. Shklyarevskii$^{1,}$ $^{2}$,  
H. van Kempen$^{1}$\\
$^{1}$ Research Institute for Materials, University of 
Nijmegen,\\ Toernooiveld 1, 6525 ED, Nijmegen, The Netherlands\\
$^{2}$ B. Verkin Institute for Low  Temperature Physics \& 
Engineering,\\ National Academy of Science of Ukraine, 47 Lenin Av.,
310164, Kharkov, Ukraine


\begin{references}



\bibitem{Keijsers95} R.J.P. Keijsers, O.I. Shklyarevskii, and H. van Kempen, 
                  \prb {\bf 51}, 5628 (1995).
\bibitem{VlaZa} K. Vladar and A. Zawadowski, \prb {\bf 28}, 1564 (1983);
		{\bf 28}, 1582 (1983); {\bf 28}, 1596 (1983).


\bibitem{KozKul} V.I. Kozub, Fiz. Tverd. Tela (Leningrad) {\bf 26}, 1995 (1984) 
[Sov. Phys. Solid State {\bf 26}, 1186 (1984)],  V.I. Kozub and I.O. Kulik, Zh. Eksp. Teor. Fiz. {\bf 91},
                 2243 (1986) [Sov. Phys. JETP {\bf 64}, 1332 (1986)]..


\bibitem{Keijsers}  R.J.P. Keijsers, O.I. Shklyarevskii, and H. van Kempen,
Phys. Rev. Lett. {\bf 77}, 3411 (1996). 
\bibitem{vonDelft} Jan von Delft, G.\ Zar\'and, and 
A.\ Zawadowski, Preceding Comment, Phys. Rev. Lett.   
\bibitem{Balkashin} O.P. Balkashin, R.J.P. Keijsers,
        H. van Kempen,  Yu.A. Kolesnichenko, and
         O.I.~Shklyarevskii, to be published.
\end{references}
\end{document}